\documentclass[]{aa}
\input epsfig.sty

\def\edoc{\end{document}}

\def\odf#1#2{\frac{{{\rm d}\kern.05em}#1}{{{\rm d}\kern.05em}#2}}

\def\L3{{}^{(3)}\!\Delta}
\def\uu{u}

\def\Y{\Theta}

\def\tsigma{\Sigma}
\def\Lorentz{{\left({1\over\density} 
	(H^i H_j -{1\over2} H^2\delta^{i}_{j})^{;j}
	      \right)_{;i}}}

\def\density{{\rho}}

\def\rw{_{(0)}}

\def\AA{{\it 3}}

\def\xx{{\bf x}}

\def\kk{{\bf k}}

\def\uv{{\uu}_{\kk(\vartheta)}}

\def\uvdust{{\uu}_{\kk(\vartheta)}}

\renewcommand{\thesection}{\arabic{section}}
\renewcommand{\thesubsection}{\arabic{section} \arabic{subsection}}

\begin{document}

\title{Spatial energy spectrum of primordial magnetic fields}

\author{Grazyna Siemieniec-Ozieblo}

\institute{Astronomical Observatory, Jagellonian University\\
Faculty of Mathematics, Physics and Computer Science\\
ul. Orla 171, 30--244 Krak\'ow, Poland}
\titlerunning{Spatial spectrum of primordial magnetic fields}
\authorrunning{G. Siemieniec-Ozieblo}

\abstract{Here, we analyze the primordial magnetic field 
transition between a radiative and a matter-dominated universe. 
The gravitational structure formation affects its evolution 
and energy spectrum. The structure excitation can trigger 
magnetic field amplification and the steepening of its 
energy density spectrum.}

\maketitle


\section{Introduction}

The increasing empirical 
evidence of magnetic fields at different cosmological scales raises a series 
of questions on their origin and their dynamical impact on gravitational 
structures. Different models of magnetogenesis have been proposed (Berezhiani \& Dolgov 2004,  
Grasso \& Rubinstein 2000, Hogan 1983)
and the possible effects on structure evolution have also been discussed (e.g. Kim et al. 1996, Wasserman 1978). 
The role of the feedback process (i.e. structure formation on magnetic field evolution) 
is equally important. 
It is vital to understand whether the structure origin and its subsequent evolution 
may change the initial spectrum of the energy density of the primordial magnetic field. 
In particular, does the structure appearance `define' the ultimate shape of the 
magnetic energy spectrum? 
In this paper we refer to the model of the structure origin, 
presented in Siemieniec \& Woszczyna 2004a (hereafter AI). It shows the existence of `explosively' growing structures at decoupling, 
originated in the stochastic field of acoustic-type perturbations in the radiation 
era.
The present paper deals primarily with the problem of the primordial magnetic field transfer 
at decoupling and its later linear evolution. The considered initial magnetic field 
is assumed to be generated in an earlier magnetogenesis process. Its final strength, 
the amplification rate at decoupling and the subsequent energy transfer to large and/or 
small scales depend on the velocity field. The presence of velocity flows at the phase 
of decoupling is a crucial aspect in the evolution of the primordial magnetic fields. 
The magnetic energy may increase at the expense of the kinetic energy of the flows,  
depending on their specific nature (e.g. shear).
 
The structure of the paper is following. In Sect. 2 we discuss the assumptions 
and introduce a set of magnetohydrodynamics equations (MHD). In Sect. 3 the solution 
of the induction equation is given. We discuss the magnetic energy density spectrum after 
decoupling in the last section.

\section{Preliminary statements and assumptions}

The main purpose of this paper is the investigation of the primordial field 
development at decoupling and directly thereafter. Assuming that 
magnetic fields were created in the early universe by some particle physics processes 
and that the fields were not correlated with the density fluctuations, the 
following questions are raised. Can  
the correlation between the magnetic field and matter density growth be established 
at recombination, i.e. if initially we have a random distribution of magnetic energy, 
then, does it become more organized later. The second one is how the final energy distribution 
is determined by the initial magnetic spectrum. The motivation for these questions 
is that the velocity which exists at recombination affects the magnetic field through 
the induction equation. Thus if structure formation develops hierarchically, 
does it mean that the cosmological magnetic field also amplifies as a bottom-up 
phenomenon resulting in an `inverse cascade' type of magnetic energy transfer? 
Even in the linear regime it seems possible that the velocity field will amplify magnetic fields 
soon after recombination and boost the magnetic energy toward larger scales.

\subsection{Basic equations and assumptions}

To study this we need to consider the magnetic evolution described by the 
induction equation, which is a part of the MHD set of equations. In some cases 
the symmetries allow its separation from the MHD system. An example of the evolution of  
selfgravitating magnetized structure with planar symmetry (pancake) recently has been 
analyzed in Siemieniec \& Woszczyna (2004b MCW). Below, we apply the full set of  
pressureless MHD equations (corresponding to eqs. (1)-(3) in MCW)

\begin{eqnarray}
\label{eq1}
\dot{\density} &=& - \density \vartheta\\
\label{eq2}
\dot{\vartheta} &=& -{1 \over 3}\vartheta^2 -{4\pi}\kappa\density - 2\sigma^2+ 2\omega^2 + \nonumber \\
                & & \Lorentz \\
\label{eq3}
\dot{H^i}&=&\sigma^i_{j}H^j+\omega^i_{j}H^j-{2\over3}\vartheta H^i, 
\end{eqnarray}
\noindent
where $\density, \vartheta, H$ are the density, expansion rate  
and the magnetic strength of the matter; the hydrodynamic scalars $\sigma$ and 
$\omega$ measure the shear and the rotation.

In order to study the cosmological development of the primordial magnetic field 
in the linear regime, we solve the above set of equations, assuming the following:

1. The evolution of the magnetic field is investigated in the post-recombination phase of 
   a spatially flat FRW universe.

2. We analyze the magnetic field perturbations (small with respect to density perturbations) 
   in the framework of the model (AI) describing the density perturbation transition between the 
   radiation and the matter-dominated era. Thus, we use here the already-derived quantities describing 
   the hydrodynamic evolution and structure formation which occurs at the transition.

3. The transition is instantaneous. In this oversimplified approach the recombination 
   coincides with the decoupling which is not a realistic assumption (at the same time one 
   neglects the viscosity phenomena that are present at this phase). We believe however 
   that the essential evolutionary features do not depend very much on this, i.e. the 
   proposed instability mechanism persists. 

4. Density and velocity perturbations evolve in both epochs. Matching of their values at `decoupling' is 
   performed according to the Darmois-Israel joining conditions (Darmois 1927, Hawking \& Ellis 1973, AI). The perturbations of the primordial 
   magnetic field $\delta H$ are however induced by the density and velocity field perturbations  
   $\delta \rho$, $\delta \vartheta$, and since it is attributed to the 
   structure formation process, they have their origin at decoupling. The magnetogenesis 
   mechanism producing the primordial seed field is expected to occur prior to 
   recombination i.e. created before the `initialization' of structure formation. 
   
5. The perturbations are rotationless but the shear introduced by collapse is important. 
   It accounts for the results. For compatibility with our earlier paper (MCW), 
   we postulate the symmetry in shear tensor of the form 
   $\sigma_{j}^i H^j = \alpha \delta \vartheta \delta_{j}^i H^j$. (Referring to MCW, 
   where the fluid contraction is one-dimensional, it means 
   that  $\alpha = {2\over 3}$ for the field component which is not contracted 
   (i.e. parallel to the velocity) and  $\alpha = - {1\over 3}$ for both components lying in the 
   pancake plane. As a result of this symmetry the shear scalar can be uniquely expressed 
   by the velocity perturbation in the induction equation. Its first term 
   is equal to $\sigma_{j}^i H^j = {\delta \vartheta \over {3}} (3\delta_i^\AA\delta_j^\AA - \delta_{ij}) H^j$.  
   Thus, if the perturbation $\delta \vartheta $ is known, then the induction equation can be 
   separated from the whole MHD system.
   
6. The spectral distributions of the hydrodynamic quantities (velocity and density fields) 
   are determined by the unknown spectrum of the primordial density perturbations  $P(k)$. 
   The primordial magnetic field spectrum is also unknown. Both come from the early 
   universe processes and are supposed to be generated independently. In our calculations we analyze 
   the scale-free spectra as test forms.

\subsection{Propagation of the linear perturbations}

The density and velocity contrasts $\delta = {\delta\rho\over \rho_{(0)}}$,  $\Theta = {\delta\vartheta\over \vartheta_{(0)}}$ calculated in (AI) are `functionals' of the primordial density 
spectrum $P(k)$. In the linear regime they are expressed by Fourier integrals 
(cf. eq. (38) i (45) in AI), where the Fourier coefficients $\{{\cal A}_{\kk}\}$ are 
 functions of the primordial initial spectrum, $|{\cal A}_{\kk}|^2 \propto P(k)$,  
and the density and velocity modes in the matter-dominated era are functions of 
conformal time $\eta$ and the wave number $k$ given by eqs. (32), (47) in AI. The linearized 
hydrodynamic evolution, eqs. (1) and (2), after introducing the perturbations, 
\begin{eqnarray}
	\label{eq4}
 \rho = \rho_{(0)} + \delta \rho\\
 	\label{eq5}
 	\vartheta = \vartheta_{(0)} + \delta \vartheta\\
 	\label{eq6}
 	  H^i = H^i_{(0)} + \delta H^i
\nonumber
\end{eqnarray}
is governed by 
\begin{eqnarray}
\label{eq7}
\dot{\delta\density}& =& -\vartheta\rw \delta\density-\density\rw \delta\vartheta\\
\label{eq8}
\dot{\delta\vartheta}&=& -{4\pi}\kappa\delta\density-{2 \over 3}\vartheta\rw\delta\vartheta, 
\end{eqnarray}
where the solutions for contrasts (${\delta\rho\over \rho_{(0)}},  {\delta\vartheta\over \vartheta_{(0)}}$) 
are given by eqs. (38), (45) in AI. The equations for background dynamics are 
\begin{eqnarray}
\dot{\density}\rw &=& \density\rw \vartheta\rw,\\
\label{eq10}
\dot{\vartheta}\rw &=& -{1 \over 3}\vartheta\rw^2 -{4\pi}\kappa\density\rw\ ,  
\end{eqnarray}
while the linearized induction equation (3) for $\delta H^i$, according to assumption 5, becomes 
\begin{equation}
\label{eq10a}
\dot{(\delta H^i)} = \left(\alpha - \frac{2}{3}\right) \, \delta\vartheta H^i_{\rm (0)} - \frac23
\vartheta_{\rm (0)} \delta H^i 
\end{equation}
and the seed field satisfies
\begin{eqnarray}
\label{eq11}
\dot{H}_{(0)} = -{2\over3}\vartheta_{(0)} H^i_{(0)} \Rightarrow H_{(0)}\propto a^{-2}
\end{eqnarray}
It is worth noticing from the above that in the case of no asymmetric contractions 
i.e. $\beta = \alpha - {2\over 3} = 0$, the induction equation leads to the known 
result (identical to the background field $H\rw$ evolution), $\delta H^i \propto a^{-2}$. 
Contrary to that, if $\beta \neq 0$, (for contraction $\beta < 0$) then in eq. (10) we have a source 
$\propto \delta\vartheta H^i\rw$, which determines the evolutionary behaviour of $\delta H^i$. 
This term itself depends on both the initial injection seed field $H\rw$ spectrum and the spectrum 
of $\delta\vartheta$. 
While the former may be postulated as a free parameter, the spectrum $\delta\vartheta$ 
is not only a product of the primordial density fluctuation spectrum but also 
a result of changes occurring at the radiation to dust transition.

The primary result of the transition model (AI) applied here is structure 
excitation with the features of acoustic shock formation at decoupling.
The changes in density and velocity field (due to this transition, see fig. 2 and 3 in AI) 
are felt by the magnetic field. Thus one can expect that the induced magnetic field 
$\delta H$ should be substantially correlated with the structure. 

\section{Induction equation}
Let us consider the propagation of the magnetic perturbations (10). For convenience 
we use here the conformal time $\eta = \int a(t)^{-1} dt$, where the scale factor \footnote{This form of $a(t)$ is a result 
of the joining conditions at the time of transition with the radiation scale factor $a(\eta)=\sqrt{\frac{\cal M}{3}}\,\eta$.}  
$a(\eta)$ has the form (eq.(21) in AI) 
$a(\eta)=\frac{\sqrt{\cal M}(\eta
+\eta_{\tsigma})^2}{4\sqrt3\eta_{\tsigma}}$, 
$\cal M$ is the constant of motion in the Friedmann equation and $\eta_{\tsigma}$ 
represents the transition time from a radiation to a matter-dominated era,  
$\vartheta_0 =3 {da/dt\over a}$. 
Equation (10) in the variable $\eta$ can be replaced by 
\begin{eqnarray}
\label{eq12}
{(\eta+\eta_\Sigma) \over 6}(\delta H^i)^\prime = \beta{\delta \vartheta \over 
\vartheta_0} H^i \rw -{2\over3}\delta H^i.
\end{eqnarray}
To obtain its solution in the linear regime we express the primordial and perturbed 
magnetic field as well as the velocity perturbation $\delta\vartheta $ as Fourier series: 
\begin{eqnarray}
\label{eq13}
H_0 &=& \eta^{-2}_\Sigma (1+\tau)^{-4} \int {\rm d}\kk e^{i\kk\xx}f(k) + \mbox{c.c.} \\
\label{eq14}
\delta H &=& \eta^{-2}_\Sigma \int {\rm d}\kk e^{i\kk\xx}G(k,\tau) + \mbox{c.c.}
\end{eqnarray}
and according to eq. (45) in (AI)
\begin{eqnarray}
\label{eq15}
\Theta \equiv {\delta\vartheta \over \vartheta_0} = \int {\rm d}\kk e^{ik\xx} \tilde{\Theta} (\tau,k) + \mbox{c.c.}
\end{eqnarray}
\noindent Equation (\ref{eq12}) becomes now 
\begin{eqnarray}
\label{eq16}
\odf{G(k,\tau)} {\tau} + {4\over \tau} G(k,\tau) = g(k,\tau) ,
\end{eqnarray}
where $\tau = (1+\frac{\eta}{\eta_{\tsigma}}) $ and 
\begin{eqnarray}
\label{eq17}
\lefteqn{g(k,\tau) = {24\pi\beta \over \tau^5} \times} \nonumber \\
  & & \!\!\!\! \int^{+\infty}_{-\infty}{\rm d}k^\prime k^{\prime2}(f(k-k^\prime)
\tilde{\Theta}(k^\prime,\tau) + f(k+k^\prime)
\tilde{\Theta}^*(k^\prime,\tau)) . 
\end{eqnarray}
The nonhomogeneous term $g(k, \tau)$ in the induction equations (16) shows the interaction 
of the mode $H_{(0)k}$ of the primordial field with the velocity modes $\Theta _k$. This term 
vanishes when $\beta = 0$. In general, the nonvanishing source of eq. (16) gives the solution 
in the form 
\begin{eqnarray}
\label{eq18}
G(k,\tau) = \left({\tau\over 2}\right)^{-4}C(k) + \left({\tau\over 2}\right)^{-4} 
\int^\tau_2 g(k,\tau)\left({\tau\over 2}\right)^4{\rm d}\tau,
\end{eqnarray}
where $C(k)$ is to be found from the boundary condition at decoupling i.e. $\delta H (\tau = 2) = 0$. 
It is seen from the above that the source modifies both the temporal and spatial (spectral) 
evolution of the primordial field.

\subsection{Comments on spectral forms}

The velocity field represented in this calculation by $\Theta$ 
is present at the decoupling time $\eta = \eta_{\tsigma}$. Together with the density contrast 
it propagates from the radiation era. Its evolution is determined by the primordial 
density fluctuation spectrum. The expansion contrast $\Theta$ is defined in the dust era 
by the joining conditions at $\eta_{\tsigma}$ and by the input parameter $P(k)$, i.e.
the energy spectrum of the acoustic field. According to AI, the general solution for $\Theta$ 
is expressed by eq. (45)
	\begin{equation}
	\label{eq19}
\Y(\eta ,\xx)
=\int({\cal A}_{\kk} \uv (\eta,\xx)
+{\cal A}_{\kk}^* \uv ^*(\eta,\xx)){\rm d}\kk,
	\end{equation} 
where the Fourier coefficients ${\cal A}_{\kk}{\cal A}_{\kk}^* \propto P(k)$  
and the modes $\uv $ after transition are given by (eq. 47 in AI)
\begin{eqnarray}
	\label{eq20}
\uvdust
&=&\frac{3^{1/4}}{\sqrt{2 k}}e^{i\kk \xx-{ik\Sigma}}\times \nonumber \\
& & \left(\frac{ik{\Sigma}}{40}
\tau ^2 + \left(4+\frac{4}{ik\Sigma} + \frac{6 ik\Sigma}{5}\right)
\tau^{-3}\right) ,
\end{eqnarray}
where $\Sigma= \eta_{\tsigma}/\sqrt{3}$ .
The comparison of Fourier expansion (15) with the above allows one to obtain the transform 
$\tilde{\Theta}(\tau, k)$. For technical reasons we take in $\tilde{\Theta}(\tau, k)$ 
only the leading term $\propto \tau^2$ (we drop $\tau^{-3}$), which is equivalent to the asymptotic 
form of the solution i.e. to a late time after recombination. For $\tilde{\Theta}(\tau, k)$ one 
finally gets 
\begin{eqnarray}
\label{eq21}
\tilde{\Theta}(k,\tau) = i\,C_1k^{1\over2}\tau^2e^{-ik\Sigma},
\end{eqnarray}
where $C_1 \equiv C {3^{1\over2}\Sigma \over 2^{3\over2}40}$ .\\

The constant C above comes from the primordial fluctuations. Both 
quantities forming the convolution in magnetic perturbation source function (\ref{eq17}) 
are two `input parameters' depending on the primordial density fluctuation spectrum $P(k)$ 
and the spectrum of the primordial magnetic fields. Both are unknown. The postulates 
concerning the first quantity commonly refer to the inflationary paradigm. Here 
we employ no concept for the primordial fluctuation generation. For the sake of 
mathematical simplicity we adopt the primordial density spectrum of the white-noise type i.e. 
$P(k) = const$ which means ${\cal A}_{\kk} = C$. For the initial primordial field, 
we do not indicate any particular magnetogenesis process with its 
specific initial spectrum . Below we assume that the spectrum generated by potentially 
different models of the early magnetogenesis can be described by the power law 
\begin{eqnarray}
\label{eq22}
f(k) &=& Ak^n	\;\;\;\; k_{\rm min} < k < k_{\rm max}\\
\nonumber f(k) &=& 0 \;\;\;\;\;\;\mbox{elsewhere}
\end{eqnarray}
In the following calculations we restrict ourselves to the analysis of two 
cases i.e. n = 0 and n = 1. In the first case we use the flat spectrum. It is plausible 
that magnetic fields have similar strength at different scales when the initial conditions 
for them were the same at the time of magnetogenesis. Thus, similar to density fluctuations, 
the white-noise spectrum is suitable. This one and the second putative spectrum,  
$f(k) \propto k^1$, allow us to put the lower spectral limit at $k_{min} = 0$. The  
spectral form used indicates that the magnetogenesis scenario occured after inflation.

\subsection{The solution}

For the above postulated categories of the spectrum, the expression (\ref{eq17}) 
becomes
\begin{eqnarray}
\label{eq23}
g(k,\tau) = \Gamma \, \bar{g}(k) \, \tau^{-3},
\end{eqnarray}
where
\begin{eqnarray*}
 \Gamma &=& 24\pi A\,C_1\beta ,\\
 \bar{g}(k)& =& i\, I_n(k) ,\\
I_n(k) &=& \int\limits^{k_{\rm max}}_{k_{\rm min}} 
{\rm d}k'((k-k')^\frac{5}{2} - 
(k' - k)^\frac{5}{2})
k'^{n}\exp({i(k' - k)\Sigma}).
\end{eqnarray*}
Finally, taking into consideration the initial condition $\delta H (\tau = 2)$
one obtains the solution for $G(k, \tau)$
\begin{eqnarray}
\label{eq24}
G(k,\tau) = {1\over2} \Gamma \, \bar{g}(k) (\tau^{-2} - 4\tau^{-4}).
\end{eqnarray}
It is important to note the structure of this solution. The temporal dependence 
of magnetic perturbations includes the typical solution of the homogeneous equation 
($G \propto \tau^{-4} \propto a^{-2}$) and the additional component $\propto \tau^{-2} \propto a^{-1}$.
Thus the sheared components, $\beta \neq 0$, of the induced magnetic fields 
are amplified in the linear regime much faster than those which are not contracted.
The spatial dependence is determined by $I_n(k)$ i.e. by the convolution of a primordial magnetic 
spectrum with the kinematic quantity $\tilde{\Theta}(\tau, k)$. 

\section{Magnetic energy density spectrum}

The evolutionary development of the magnetic field after recombination with no source term is well 
known. In the absence of velocity fields its adiabatic decrease occurs during 
the expansion of the universe (e.g. Rees \& Reinhardt 1972). The situation looks different if there is 
shear motion. In the considered model the gravitational structure begins to 
form at the recombination time. According to the solution (\ref{eq24}), the magnitude 
of the magnetic strength may substantially increase. The question arises whether  
generation of the large scale field is possible and what is the asymptotic form of its 
spectrum.

The Fourier mode of the total field is 
\begin{eqnarray}
\label{eq25}
\lefteqn{H_k = (H_{(0)})_k + (\delta H)_k =} \nonumber \\
& & \eta_{\Sigma}^{-2}\tau^{-4}f(k) + \eta_{\Sigma}^{-2}
\, \bar{\Gamma} \, \bar{g}(k)(\tau^{-2} - 4\tau^{-4}) .
\end{eqnarray}
Then in calculating of the magnetic energy associated with the mode $H_k$, we restrict ourselves 
to the leading term in $\tau$, obtaining  
\begin{eqnarray}
\label{eq26}
|H_k|^2 = {1\over4} \eta_{\Sigma}^{-4} \, {\bar\Gamma}^2 \, \tau^{-4} |I_n(k)|^2 .
\end{eqnarray}
The resulting spectrum evolves through interaction between the velocity and 
magnetic modes which changes its initial form. The tendency of these changes 
is depicted by verification of two particular magnetic spectral forms. The 
analytical calculation has been performed for $|I_n(k)|^2$, (n = 0, 1). The resultant 
energy density spectra are presented below. For a given value of parameters $k_{max}, \Sigma$ 
and for $k_{min} = 0$ we show in Fig. 1 and 2 the k-dependence of $|I_n(k)|^2$ - 
the quantity proportional to the magnetic energy density of the mode $H_k$. 
Two distinct features emerge: an energetic discrimination of $|I_n(k)|^2$ 
as a function of $k$ and the large amplitudes of $|H(k)|^2  \propto \beta^2\, |I_n(k)|^2\sim \beta^2\, 10^{10}$. 
\begin{figure}[h]  
\centerline{\psfig{file=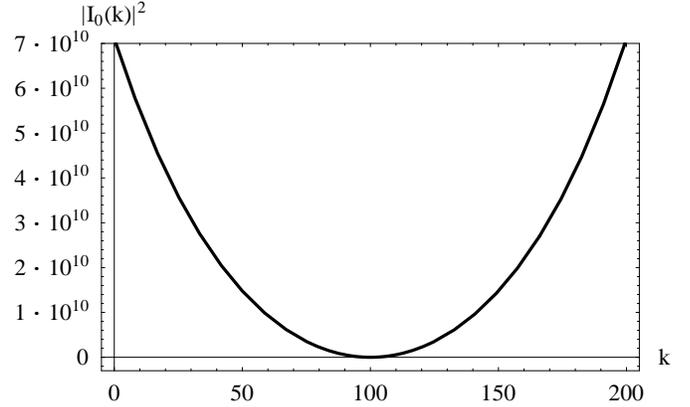,width=0.5\textwidth}\hspace{-0.5em}}
\caption{The quantity $|I_0(k)|^2$ proportional to the magnetic energy density spectrum as 
a function of the wave number.}
     \label{fig:01}  
     \end{figure}
\begin{figure}[h]
\centerline{\psfig{file=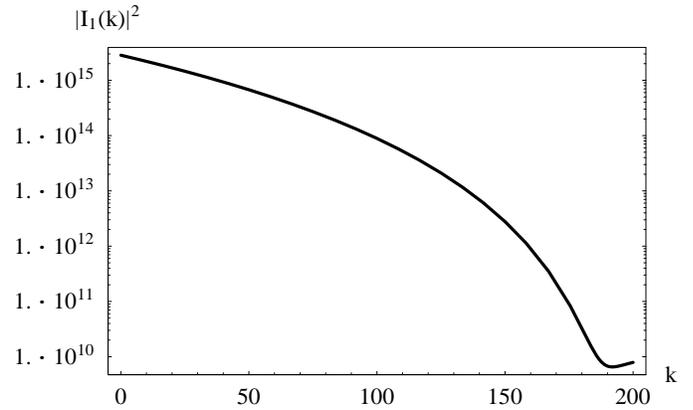,width=0.5\textwidth}\hspace{-0.5em}}
\caption{The same as above for $|I_1(k)|^2$. Axis of ordinates is in log scale.}
\label{fig:02}
	\end{figure}	
Depending on the initial magnetic spectrum, the magnetic energy is moved either 
to smaller and/or larger $k$. In the first case, there is an amplification of the 
`magnetic noise' by shearing and the energy is carried to small and large scales. For the initial power 
spectrum $\propto  f^2 \propto k^2$ the energy transfer goes backward i.e. from larger wave numbers 
to smaller wave numbers forming a sort of inverse cascade which qualitatively shows 
the results of stretching. In both cases the hydrodynamic spectrum boosts the energy to 
different wave intervals. This in turn implies a correlation between the magnetic field 
and the density/velocity fields in structures.

\section {Conclusions}

In the presented model magnetic changes result from 
the coupling of velocity shear with the primordial magnetic field. The former appears 
due to the rapid initialization of structure collapse. The direct implications are 

\begin{itemize}

\item 
field amplification and magnetic energy shift towards the larger scales 
(inverse cascading) and lead to much steeper magnetic spectra than the 
initial injection spectra. 

\item 
The shape of the final magnetic spectrum is a function 
of the initial magnetic spectrum (and implicitly depends on the primordial 
density fluctuations). 

\item 
The primordial field represents here the indispensable seed field for the future 
amplification (dynamo). Its strength at different length scales may differ by 
several orders of magnitude. 

\item 
Also a debate  about the top-down versus bottom-up scenario 
for the magnetic field origin in structures may be rather unfounded. The seed field may appear 
simultaneously at all structure scales with strongly diverse strength.
\end{itemize}

\section*{Acknowledgments}
I am grateful to Zdzislaw Golda and to Andrzej Woszczyna for discussions and for reading of the manuscript. This work is supported in part by `Komitet Bada\'n Naukowych' project `Cosmic magnetic fields'.

\def \A&A{{A\&A}}
\def \ApJ{{ApJ}}
\def \ApJS{{ApJS}}
\def \JP{{J. Phys. \rm(USSR)}}

\section*{References}   

\parskip=0pt   
\parindent=7mm
\noindent
Berezhiani, Z., \& Dolgov, A., 2004, Astropart. Phys., 21, 59\\
Darmois, G., 1927, {\em M\'{e}morial de Sciences 	Math\'{e}matiques, Fasc XXV,
                   Les Equations de la Gravitation	Einsteinienne}, (Gauthier-Villars, Paris)\\
Grasso, D., \& Rubinstein, H., 2001, Phys. Rep., 348, 163\\
Hawking, S., \& Ellis, G., 1973, {\it The Large Scale Structure of Space-Time}
                                      (Cambridge University Press, London)\\
Hogan, C., 1983, PRL, 51, 1488\\
Kim, E., Olinto, A., \& Rosner, R., 1996, \ApJ, 468, 28\\
Rees, M., \& Reinhardt, M., 1972, \A&A, 19, 189\\
Siemieniec, G., \& Woszczyna, A., 2004a, \A&A, 419, 801 \\
Siemieniec, G., \& Woszczyna, A., 2004b, \A&A, 414, 1 \\
Wasserman, I., 1978, \ApJ, 224, 337\\

\end{document}